\documentclass[10pt,a4paper]{article}

\usepackage{amsmath}
\usepackage{amsfonts}
\usepackage{amssymb}
\usepackage{graphicx}
\usepackage{color}
\usepackage{subfig}
\usepackage{url}

\title{Cyber Physical Systems: Prospects and Challenges}
\author{Walid Gomaa $^{1,2}$\\ 
\small $^1$ Cyber-Physical Systems Lab,\\
\small Egypt Japan University of Science and Technology,\\
\small Alexandria, Egypt.\\
\small $^2$ Faculty of Engineering,\\
\small  Alexandria University, Alexandria, Egypt.\\
\small walid.gomaa@ejust.edu.eg}

\begin{document}

\maketitle

\begin{abstract}
Cyber physical systems CPSs embodies the conception as well as the implementation of the integration of the state-of-art technologies in sensing, communication, computing, and control. 
Such systems incorporate new trends such as cloud computing, mobile computing, mobile sensing, new modes of communications, wearables, etc.
In this article we give an exposition of the architecture of a typical CPS system and the prospects of such systems in the development of the modern world.
We illustrate the three major challenges faced by a CPS system: the need for rigorous numerical computation, the limitation of the current wireless communication bandwidth, and the computation/storage limitation by mobility and energy consumption.
We address each one of these exposing the current techniques devised to solve each one of them.
\end{abstract}
\noindent keywords: {Cyber Physical Systems; rigorous computation; wireless communication; deep neural networks; mobile sensing; mobile computing}

\section{Introduction}

\label{sec: Introduction}

The modern world is witnessing an explosion in the computing and communication technology. Computing power and facilities are becoming more and more integrated into every aspect of
our private and public lives.
Computing devices can be found everywhere from traditional computing machines such workstations, laptops, etc, to smart phones and smart home appliances, and finally with the current revolution 
in wearable devices such as headsets (for virtual and augmented reality VR/AR), smart watches, etc. Communication networks, particularly wireless networks, have become very mundane and widespread throughout the globe.
Web connection is now very popular and affordable to a large portion of the population. The raw material for computation and communication is 
data which is typically gathered through sensing equipment, which are 
now embedded in most popular devices and appliances. Such sensors include different varieties of cameras, acceleromters, gyroscopes, barometers, weather stations, VR/AR headsets, etc. Similarly, the other way 
around of affecting the environment via actuators operated by control systems such as in robotics and different varieties of mechtronic systems. 
The integration of all of these: sensing the physical environment, data communications, computation, and actuation, forms what is called a \emph{cyber-physical system (CPS)}.
This terminology is typically used by computer and control scientists and engineers.
Another term that has been commonly used by communication scientists and engineers, though limited in its scope than CPS, is the \emph{Internet of Things (IOT)}. 
For the rest of this article we use `cyber physical systems' to mean such conception.

\begin{figure*}[h]
\centering
\includegraphics[scale=0.35]{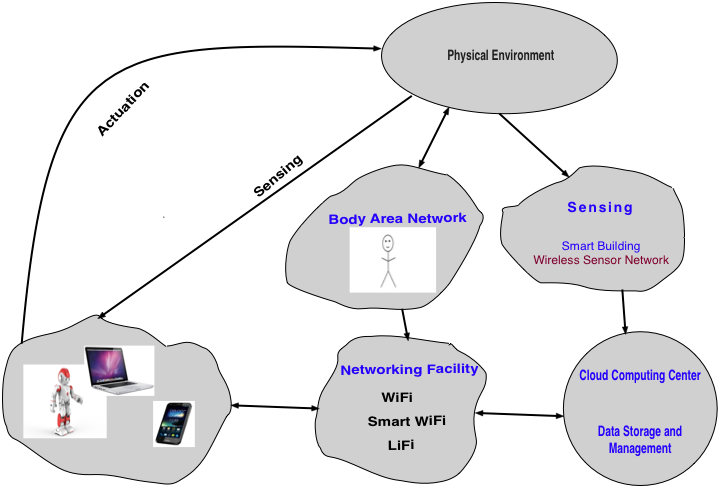}
\caption{\label{fig: CPS} A typical illustration of a CPS system}
\end{figure*}

Fig.1 illustrates the main components of a typical CPS system. We can figure out four major components: sensing, communication, computing, and actuation/control. We are currently witnessing 
a revolution in the sensing technology.
Sensors are embedded in almost every electronic device including: smart watches, home appliances, air conditioning, vehicles electronic system, computers, 
wearables, smart phones, etc. Prices are getting cheaper, sizes are getting smaller, and more quantities and qualities are being measured by sensors. Sensors enable the CPS system 
perceive the environment. Of course, there must be circuitry designed to convert the electric signals measured by the sensor into the target property such as temperature, humidity, velocity, 
acceleration, etc.
Communication provides the necessary interconnectivity among the different components and parts of the CPS system. Communication technologies can vary from Wi-Fi to wired networks to communication
through power lines, etc. The third main component is `computation', the transformation of incoming data into another space that is dependent on the underlying process and/or application.
The computing power today is shifted from the traditional Moore's law into a new form. Moore's law states that: the number of transistors on a microprocessor will double every two years or so, and 
accordingly the computing performance will. However, we have currently reached the nano-scale level in transistors manufacturing, and we are almost at the end barrier, after which we 
will hit the atom level scale and quantum effects and uncertainties will prevail making things complicated and unreliable. So the Moore's law has somehow be reinterpreted to indicate the same 
goal of exponential performance increase, however, not through increasing the number of transistors, but through other more abstract means. One advancement is to include more than 
one processor, called \emph{core}, in the chip. For example, four and eight cores are common in most of today's desktop computers, laptops, and even smartphones~\cite{waldrop_chips_2016}.
High performance is then achieved through \emph{parallelism}. The final major component of a CPS system is that of control/actuation. This can be considered as the converse of sensing.
Through actuation we want to change the current state of the environment. Examples include the thermostats of an air-conditioning system to change the temperature of the surrounding environment, a robot arm
to move an object to a new place, the legs joints of a robot to move it to a new place, etc.

\begin{figure}[h]
\centering
\includegraphics[scale=0.7]{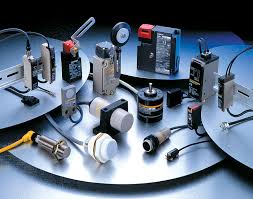}
\caption{A bundle of different sensors}
\end{figure}

In this article we discuss some essential challenges that face the design, implementation, and deployment of modern cyber-physical systems. The first challenge is concerned with the need 
for rigorous numerical computation in order to first deal with the kind of non-traditional computational requirements of such systems and second to obtain reliable computation and consequential outputs.
The second major challenge is that of the limitation of the current technology of wireless communications. The current spectrum will not be enough to cover the prospect transmission demands from 
all kinds of devices including wearbles such as headsets for virtual and augment realty which are heavily hungry for bandwidth. The third major challenge that we illustrate in the current paper is that of the computation and 
storage requirements for such new kind of computation.

Section~\ref{sec: Introduction} is an introduction. Section~\ref{sec: Non-Rigorous Computation} addresses the challenge of rigorous numerical computation. Section~\ref{sec: Communication and Networking} addresses the current status and crisis
communication and networking. Finally, section~\ref{sec: Computing/Storage Power} addresses the issues and new trends in the computation realm.

\section{Non-Rigorous Computation}

\label{sec: Non-Rigorous Computation}

The first fundamental problem encountered in CPS systems is the discrete/continuous dilemma; said another way the digital/analog dilemma. 
Modern computer and communication systems are inherently digital in nature: every object, over which computation is performed and/or needs to be transmitted over communication channels, is represented by 
a finite string.  This string consists of letters from an alphabet, typically taken to be the binary alphabet $\{0,1\}$. For example, if the object of computation is an integer, then the integer is represented in its binary radix: 
$5$ is represented by $101$, etc. A rational number such as $\frac{2}{3}$ can be represented as
$10\#11$. This goes along with any object with finitary nature: encode it with a finite binary string. Computation can then be performed exactly, there is no loss of information and/or accuracy.
However, the problem arises when we deal with real physical and engineered systems, where such systems are typically designed, modeled, 
and analyzed using mathematical analysis tools whose underlying objects have inherently infinitary nature such as the real numbers, complex numbers, continuous functions, manifolds, etc.
For example, weather prediction systems are modeled using a set of three differential equations, called the Lorenz equations (with variant extensions). Mechanical systems are typically modeled 
by Lagrangian mechanics, etc. The variables and parameters of all these equations range over the real and complex numbers.

In order to appreciate the essential difficulty in representing such objects, take a very simple example as representing the real number $\sqrt{2}$. This number has no finite pattern and so no finite representation.
So it can not be represented exactly on a digital computer, there must be a loss of information and accuracy.
The best feasible solution is to approximate $\sqrt{2}$ within a finite known precision. For example, the number $1.4142$ (floating point or rational as a computer data type) is an approximation 
to $\sqrt{2}$ within an error of $10^{-4}$, that is, $|\sqrt{2} - 1.4142| \le 10^{-4}$.  Though this approximation may be good enough and controlled at the current point of execution, the error
can propagate and expand in unexpected way with further operations such as arithmetic and/or logical ones. Failure to address such discrepancy between the discrete and the continuous, 
non-rigorous computation, can lead to unimaginable disasters. In the following we cite one of them.

\begin{figure}[h]
\centering
\includegraphics{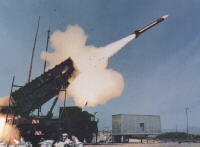}
\caption{\label{fig:patriot_missile} Patriot missile}
\end{figure}

\noindent\textbf{Patriot missile failure:} During the Gulf War on February 25th 1991, an American Patriot Missile battery in Dharan, Saudi Arabia,
failed to track and intercept an incoming Iraqi Scud missile; see fig~\ref{fig:patriot_missile}. This resulted in the killing of 28 soldiers and injuring around 100 other people 
(this is almost the whole American casualities during the entire war).
Investigation into the incident concluded that this is essentially due to a software failure, specifically an apparent case of \emph{non-rigorous computational numerics}.
The technical problem is described as follows. The time in tenths of second as measured by the system's internal clock was multiplied by $\frac{1}{10}$ to produce the time in seconds.
This calculation was performed using a $24$-bit fixed point register.
The value $\frac{1}{10}$ has an infinite binary expansion: \\
$(0.0001100110011001100110011001100....)_2$. The $24$ bit register in the Patriot rocket only stored\\ $(0.000110011001100110011001)_2$.
So the number was truncated at $24$ bits after the binary radix point introducing an error of \\$(0.0000000000000000000000011001100...)_2$, or about $0.000000095$ in decimal.
This small truncation error, when multiplied by the large number giving the time in tenths of a second, led to a significant error.
The Patriot rocket battery had been up for around $100$ hours, so multiplying by the number of tenths of a second in $100$ hours magnifies the induced error
to give $0.000000095 \times 100 \times 60 \times 60 \times 10 = 0.34$ seconds.
The attacking Scud missile travels at about $1,676$ meters per second, and so travels more than half a kilometer in $0.34$ seconds.
This was far enough that the incoming Scud was outside the range window that the Patriot tracked.

\begin{figure}[h]
\subfloat[\tiny Discretized phase space: the dotted rectangle is the exact image and the gray area is the ideal representation of $f(P_i)$)~\cite{EGIdeal}]{\includegraphics[scale=0.5]{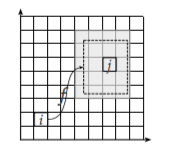}} \qquad
\subfloat[\tiny Every node $i$ maps to itself due to the original map behavior, but maps to elements $i − 1$ and $i + 1$ also due to overlapping only.]{\includegraphics[scale=0.45]{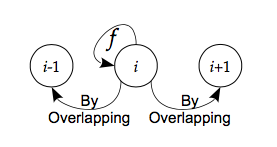}}
\caption{\label{fig: rigorous simulation 1} Combinatorial representation of the logistic map}
\end{figure}

The key to the solution to this problem is \emph{rigorous computation}. It is a very wide spectrum area of research ranging from the pure theoretical to the pure practical. From the theoretical side the area of \emph{computable analysis}
investigates mathematical analysis from the computablility perspective. It provides a foundational framework for the study of computablility and complexity theoretic of what can be done
effectively and efficiently over continuous domains.
An example of a result in this field is as follows. Assume a function $f \colon [0,1] \to \mathbb{R}$. Then, $f$ is computable if and only if $f$ is effectively continuous and effectively approximable. Intuitively, this means $f$ is continuous and its continuity can 
be effectively computed and the function values can be approximated by a computable function over discrete objects (the integers or the rationals).
The books~\cite{KComplexity,WComputable} provide general and advanced introduction to the subject. Algebraic characterizations of computable analysis (an algebraic perspective of the computability and complexity
theoretic conceptions) can be found in~\cite{FGHAnalytical,GRational,BGHAlgebraic,GAlgebraic,WGSurvey}. 
The next step down the abstraction is \emph{numerical analysis}. This area provides the necessary algorithms for numerical calculations of the vast majority of problems in mathematical analysis.
For example, root finding, integration, solution of differential equations, solution of systems of linear and non-linear equations. Such plethora of algorithms come with numerical analysis techniques and tools indicating different factors such as convergence guarantee and rate, stability, error 
bounding, etc. For an introduction to the subject, see~\cite{burden_numerical_2015, hamming_numerical_1987}. 
The third step down the hierarchy of rigorous computation is providing the necessary data structures for the proper handling, both from the computability and complexity 
perspectives, of numerical computations. This is evident, for example, in the rigorous simulation of dynamical systems which can also be used for computer-assisted proofs about such systems.  A notable example of such kind of research is the framework of 
\emph{finite resolution dynamics} introduced by Stefano Luzzato et al. in~\cite{LPFinite}. They introduce a combinatorial representation of discrete-time continuous-space dynamical systems as a graph, and map mathematical properties of dynamical systems 
such as transitivity and mixing to graph-theoretic properties such as connectivity and aperiodicity. An improvement over such framework from the perspective of computational efficiency is introduced by I. Elshaarawy and W. Gomaa in~\cite{EGIdeal,EGEfficient}.
Fig~\ref{fig: rigorous simulation 1} shows an illustration of the combinatorial graph representation of the famous logistic map $x_{t+1} = rx_t(1-x_t)$ over the unit interval $[0,1]$.

\section{Communication and Networking}

\label{sec: Communication and Networking}

An essential component in a CPS/IOT system is communication network. One of the big hurdles and challenges in such systems is the data traffic jam. By the end of year 2014, the global \emph{mobile} data traffic reached about $2.5$ exabytes/month 
($2.5\times 10^{18}$ bytes) according to Cisco systems. Around $100$ million wearable devices around the world have been contributing about $15$ million gigabytes/month (about 
$0.6\%$ of the total mobile traffic) and this number is expected to increase five-fold by 2019~\cite{AWhat}. All these devices may consume and choke off the available bandwidth and hence,  degrade the communication performance of the Internet traffic. 
There are several possibilities to tackle this problem. 

The first intuitive solution is to free up an extra bandwidth, essentially in the radio spectrum, that is used, for example, in other purposes such as the military.
One implementation of this strategy is done by the United States government who pledged in 2010 to free up an extra $500$MHz which doubles the bandwidth available for mobile devices at that time~\cite{AWhat}.
However, this traditional solution is unlikely to be enough to cope with the huge proliferation and increase in the technology and number of wearable devices, particularly, headsets used
for virtual and augmented reality VR/AR applications.
Besides, this solution may lack standardization as dealing with limited bandwidth can be specific to each country according to its own regularization and governmental usage.
For example, Indaia has access to only $10\%$ of the bandwidth available to people in the USA~\cite{AWhat}. Of course, there are many other issues with the current traditional wireless radio communication including efficiency regarding energy consumption, availability (such as in hospitals, airplanes, etc), and security and privacy.
So a more demanding solution is to use the available bandwidth in a more efficient way. One possibility is to create a hierarchy of networks. Wearables on a particular person should 
communicate with each other through a local network, coined as \emph{body-area network }, which is designed to use a different part of the electromagnetic spectrum such as the millimetre wavelenngth MmWave ranging from $30$ GHz to $300$ GHz.
Body area network consists essentially of wearable devices in and around the human body. It is supposed to connect low rate sensors such as accelerometers and pedometers and high rate devices such as VR/AR headsets.
A prominent researcher in the MmWave network is Robert Heath who is the director of the Wireless Networking and Communications Group and the Wireless Systems Innovations Lab in the 
University of Texas At Austin. The group has developed theoretical models for performance analysis of such wearable networks under different situations and scenarios including the effect of reflection of walls, ceilings, and floors in addition to the effect of blocking by the particular pose and orientation of the underlying person.
Such kinds of networks might use wireless standards like IEEE 802.11ad or WirelessHD, based on which, commercial products are already available.
One device can act as a hub in the body-area network which uses the MmWave spectrum and would use congested conventional bands to communicate data to the Internet.

\begin{figure}[h]
\centering
\includegraphics[scale=0.33]{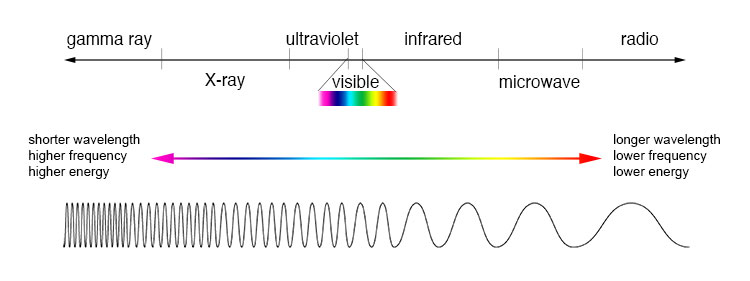}
\caption{\label{fig: electromagnetic spectrum} Electromagnetic spectrum}
\end{figure}

A rather radical approach to the communication bottleneck is the use of the available spectrum that comes from the visible light radiation from light-emitting diodes LEDS. LEDS produce light and can act also as photoreceptors. By looking at the electromagnetic spectrum as shown in fig~\ref{fig: electromagnetic spectrum}, we can have a global perspective on the potentiality for wireless communication.

The gamma rays band is dangerous health-wise. The technology of X-rays is in general not that popular, except in hospitals, and are expensive for large scale use. The ultraviolet light can also be dangerous for the human body. Infrared can only be used with low power for safety of the eyes. The radio and microwave band is the current technology used
for wireless communication. However, this current technology is facing several challenges including:

\begin{enumerate}
\item Scarcity of the available radio bandwidth.
\item Large amount of energy is needed for radio communication.
\item Due to safety and security issues radio WiFi can not be available in several situations such as hospitals, flights, etc.
\item There are many security holes in radio transmission, they can penetrate through walls and be intercepted.
\end{enumerate}

On the other hand communication via the visible light spectrum potentially provides solutions to all such problems. The radio and microwave spectrum range from several KHz up to 
several GHz, whereas the visible light ranges from the early GHz up to early PHz (Peta Hz) implying that with LEDS we have about more than $10^4$ wider spectrum, and we already 
have more than $10^4$ LEDS already installed in the infrastructure (compared with the current wireless facilities).
Therefore, we have much more capacity than the current radio wireless.  LEDS are efficient source of illumination from the perspective of energy consumption.
And at the same time they serve as data carriers which greatly improves the energy budget over the current radio transmission. Illumination can be found everywhere and LEDS can be installed and replace  other kinds of illumination at relatively cheap prices. Hence, communication availability comes for free. Data transmission through illumination has better security than radio transmission.
On one hand, light does not penetrate through walls; on the other hand light sources can be easily moved away from the receptors in order not to receive data.

This wireless technology based on the light wave spectrum is prospected to be used for both communication among the wearables as well as communication with the global Internet. Body-area network on a person can communicate through wearables that incorporate LEDS for such kind of communication. 
Furthermore, such information would be sent to light installations in the location (an office for example), that would be connected to the Internet through
the \emph{power wiring}. An important thing to mention here is that blinking of LEDS for data transmissions is so fast that it is imperceptible to the human eyes.
One of the leading research groups in the world in this area is the mobile communications group at the University of Edinburgh, UK, led by Professor Harald Haas.
They call this new technology Li-Fi. Harald Haas plans to test a Li-Fi system in hospitals this year 2016, where patients will wear wristbands that monitor their temperature and transmit the
data using LEDS on these wristbands to the hosptial's lighting system~\cite{AWhat}.

Another way to overcome the wireless communication hurdle is the smart use of the current technology by reducing the amount of redundant data.
The idea is similar in abstraction to Li-Fi in that multi-tiered network is used.
A kind of a local network, for example, among people in the same area trying to access the same information content from the Internet such as that of a sports match or
a musical concert: one device can act as a hub seeding the same information content to all other devices around through the local network.
Actually, this is the idea of the fifth generation 5G mobile communication systems.

\section{Computing/Storage Power}

\label{sec: Computing/Storage Power}

Generally, there are two trends regarding the processing of the huge amount of stream data available today. The first one is \emph{cloud computing},
where the mobile devices (and/or fixed devices), whatever they are,
transmit their data to a cloud computing center, where all the software tools and processing are done, and then results are relied back to the mobile devices.
Cloud computing has high performance computing power as well as massive storage capabilities and so relieves the burden of computing and storage from the mobile devices.
However, this creates a pressure on the limited communication resources as mentioned in the previous section.
A notable example of the use of the cloud computing is the area of \emph{cloud robotics}.
Unlike the traditional approach to robotics systems where all sensing, computing, and memory 
are integrated into a single standalone system, computing and memory storage are relied to cloud computing using the additional resource of communication networks.
A preliminary conception of the idea can be found at~\cite{IRemote,IKKHIPlatform}. An active research group in this area is the Jouhou System Kougaku Lab at the University of 
Tokyo in Japan \url{http://www.jsk.t.u-tokyo.ac.jp/research/rbr/portablehumanoid.html}. And a survey of the current research in cloud robotics can be found at~\cite{KPAKSurvey}.

The second trend is to use on-device CPU and GPU capabilities (embedded CPU/GPU processors) in a smart way in order to perform quick and critical computations without the need of a remote computing power that maybe 
blocked by the limitation of the communication channel. This creates a new area of research and application that is concerned with the integration of mobile sensing and mobile computing. This is yet to be investigated, particularly, the use
of the machine learning technology in object recognition, speech recognition, activity recognition, image classification, etc. Although mobile sensing/computing share many of the same data modeling challenges as in other arenas, there are unique characteristics in the mobile case.
This involves the following:

\begin{enumerate}
\item Measurements of mobile sensors can be greatly  affected by the particular location, orientation, and context of the embedding mobile device. For example, a smartphone can be in a pocket, a bag, or in a conversation context, etc.
\item Mobile sensors are highly noisy, either by the inherent noise of the physical characteristics of the sensors, or by the background noise that can vary significantly according to the particular context and situation such as driving, resting, indoors, outdoors, etc.
Modeling of such noise processes can be very complicated.
\item Unlike fixed sensing devices, mobile sensors are by nature very personalized. Measurements can depend to a large extent on the particular user and her style of living.
\item Power consumption should be an essential issue of any mobile application.
\end{enumerate}

All the above challenges can be addressed by the machine learning technology.
Deep learning in particular has been investigated for mobile sensing/computing.  On one hand this is mainly due to the fact that such paradigm has achieved the state-of-the-art accuracy in many tasks related to speech recognition and computer vision 
such as object detection, recognition, and segmentation which are the main ingredients for higher level more complicated tasks such as activity, behavior, and emotion recognition, and speaker recognition. 
On the other hand, mobile computing architectures have advanced much in recent years alleviating many of the computational constraints in previous generations of mobile devices. For example, iPhone 6 is a 10x computational improvement
over the previous iPhone 3GS~\cite{lane_can_2015}.
The main challenge with deep learning networks are the tremendous amount of computational and storage resources required for training and testing. For example, AlexNet~\cite{NIPS2012_4824}, a moderate size 
convolution deep network uses $2.3$ million weights making up about $4.6$MB of storage and correspondingly requires huge processing power.
The large number of weights and channels require substantial data movement which consumes a great deal of energy~\cite{chen_eyeriss:_2016}.
These issues become more pressing when migrating over mobile computing. Some recent work have addressed deep learning on embedded devices, however, there is still much to do. In the following we list two prominent research in that direction.

The authors in~\cite{lane_can_2015} design and implement a low-power deep neural network inference engine that uses both the CPU and DSP (Digital Signal Processing) capabilities 
of mobile devices such as the smartphones Samsung Galaxy S5 and Nexus 6.
Their prototype consists of two phases. An offline phase for training which is performed using deep neural network on conventional computing, 
and an online realtime phase that is operating on the mobile device.
They test their proposed framework against other common modeling techniques such as decision trees and support vector machines using datasets for common behavioral inference tasks. These involve
several modes of activity recognition, emotion recognition, and speaker identification. They achieve better or comparable results from the accuracy perspective and at the same time they use 
much lower resources (lower energy and runtime overhead) than the conventional Gaussian mixture models and comparable resources with the decision tree models.

The authors in~\cite{lane_can_2015} have used a simple version of deep neural networks. However, in order to achieve the state-of-the-art accuracy for typical \emph{vision} tasks such as object 
detection, recognition, and segmentation, convolutional neural networks CNN should be adapted.
CNNs are increasingly being used; the only problem with them is the huge amount of computational 
resources required by an operational CNN (such as the AlexNet network mentioned above). 
Y. Chen, T. Krishna, J. Emer, and V. Sze~\cite{chen_eyeriss:_2016} have developed a hardware solution to the problem by designing and implementing an accelerator for CNN that maintains the state-of-the-art accuracy whilst minimizing the energy consumption in the system in real-time. Their design is based on two key 
methods: (1) efficient dataflow (which in general accounts for a great deal of the consumed energy) along with the supporting hardware including spatial array, memory hierarchy, and on-chip network; the aim
is to minimize dataflow through data reuse and the support of different shapes (of the different layers of the network) and (2) exploitation of data statistics in order to minimize energy consumption through avoiding unnecessary reads and computations.
They implemented a test chip for the AlexNet convolutional neural network.  They characterize the details of their design and implementation along with the 
energy consumption at every layer of the network and every component of the chip. Briefly, they can achieve a frame rate of $34.7$fps at a power of $278$mW at $1$V.  They can increase the frame rate to $44.8$fps at the expense of an increase on the voltage limit to $1.17$V. 
The authors gave a very well detailed charts about the relationship between the performance measured in terms of frame rate
and the corresponding energy efficiency. They also provide a performance chart of the different five layers of the AlexNet CNN in terms of the reduction of data movement (DRAM access).

\bibliographystyle{abbrv}
\bibliography{references}

\end{document}